%% file: main.tex
\documentclass[sigconf,nonacm]{acmart}

\input{preamble.tex}
\begin{document}

\title{
    NOVA: Coordinated Test Selection and Bayes-Optimized Constrained Randomization for Accelerated Coverage Closure
}

\author{Weijie Peng}
\email{weijiepeng@pku.edu.cn}
\affiliation{
    \institution{Peking University}
    \city{Beijing}
    \country{China}}

\author{Nanbing Li}
\email{2501210603@stu.pku.edu.cn}
\affiliation{
    \institution{Peking University}
    \city{Beijing}
    \country{China}}

\author{Jin Luo}
\email{luo-jin@pku.edu.cn}
\affiliation{
    \institution{Peking University}
    \city{Beijing}
    \country{China}}

\author{Shuai Wang}
\email{wangshuai@primarius-tech.com}
\affiliation{
    \institution{Primarius Technologies Co., Ltd.}
    \city{Shanghai}
    \country{China}}

\author{Yihui Li}
\email{liyh@primarius-tech.com}
\affiliation{
    \institution{Primarius Technologies Co., Ltd.}
    \city{Shanghai}
    \country{China}}

\author{Jun Fang}
\email{fangjun@primarius-tech.com}
\affiliation{
    \institution{Primarius Technologies Co., Ltd.}
    \city{Shanghai}
    \country{China}}

\author{Yun (Eric) Liang}
\email{ericlyun@pku.edu.cn}
\affiliation{
    \institution{Peking University}
    \city{Beijing}
    \country{China}}

\input{sec/0_abs.tex}

\maketitle

\input{sec/1_intro.tex}
\input{sec/2_background.tex}
\input{sec/3_overview.tex}

\input{sec/4_method.tex}
\input{sec/5_eval.tex}

\input{sec/7_conclusion.tex}

\bibliographystyle{ACM-Reference-Format}
\bibliography{refs.bib}

\end{document}

%% file: preamble.tex
\usepackage[linesnumbered,ruled,vlined]{algorithm2e}
\usepackage{subcaption}
\usepackage{xspace}
\usepackage{tikz}
\usepackage{soul}
\usepackage{tabularray}

\usepackage{hyperref}
\usepackage[ruled,vlined]{algorithm2e}

\definecolor{myorange}{RGB}{245,133,24}
\definecolor{myred}{RGB}{228,87,86}
\definecolor{myblue}{RGB}{76,120,168}
\definecolor{mygreen}{RGB}{44,160,44}
\newcommand\shortline{{\rule[0.5ex]{2em}{0.8pt}}}
\newcommand\nova{{\sffamily\small NOVA}\xspace}
\newcommand*\circled[1]{\tikz[baseline=(char.base)]{
            \node[shape=circle,fill,inner sep=1pt] (char) {\textcolor{white}{#1}};}}

\setcopyright{acmlicensed}
\copyrightyear{2025}
\acmYear{2025}
\acmDOI{XXXXXXX.XXXXXXX}

%% file: sec/0_abs.tex
\begin{abstract}
Functional verification relies on large simulation-based regressions. Traditional test selection relies on static test features and overlooks actual coverage behavior, wasting substantial simulation time, while constrained random stimuli generation depends on manually crafted distributions that are difficult to design and often ineffective.
We present \nova, a framework that coordinates coverage-aware test selection with Bayes-optimized constrained randomization.
\nova extracts fine-grained coverage features to filter redundant tests and modifies the constraint solver to expose parameterized decision strategies whose settings are tuned via Bayesian optimization to maximize coverage growth.
Across multiple RTL designs, \nova achieves up to a 2.82$\times$ coverage convergence speedup without requiring human-crafted heuristics.
\end{abstract}

%% file: sec/1_intro.tex
\section{Introduction}

Simulation-based functional verification is the most widely adopted approach for validating hardware designs. Verification engineers construct RTL testbenches to generate input stimuli from given constraints for the design under test (DUT)~\cite{spear2006systemverilog}, with the goal of achieving comprehensive functional coverage~\cite{coverage, trend2016} while minimizing simulation time and computational resources.
However, RTL verification often requires simulating thousands of tests to capture potential edge cases, making exhaustive simulation prohibitively expensive.
Achieving high coverage efficiently hinges on two fundamental challenges: selecting tests that maximize coverage without redundantly exercising the same functional points, and generating stimuli that effectively target unexplored areas.

Prior test selection approaches~\cite{guzey_dac08, novelty10, novelty12, vae_dvcon21, selection_aspdac_22, ats10, nvidia_dac23} are fundamentally coverage-agnostic: they select tests based on diversity in static feature spaces, assuming that dissimilar tests will hit different coverage points.
This assumption breaks down in late-stage verification, when most common behaviors are already covered and only a small fraction of the coverage space remains untested~\cite{cdg_todaes_survey}.
In this sparse regime, feature-space diversity no longer correlates with coverage improvement. Tests may be highly dissimilar yet exercise already-covered behaviors, while similar tests might differ only in specific parameters needed to trigger rare corner cases.
Without explicit guidance from the coverage model, these approaches cannot effectively target uncovered regions.

Beyond test selection, generating high-quality stimuli is also critical.
Industrial verification workflows rely predominantly on constrained random verification (CRV)~\cite{spear2006systemverilog}, where constraint solvers generate valid stimuli for SystemVerilog UVM ~\cite{ieee_uvm} testbenches.
CRV has become the de facto standard due to its compatibility with existing infrastructure and its ability to express complex input constraints declaratively.
However, CRV solvers operate as black boxes: their internal branching heuristics and randomization strategies cannot be easily adjusted to target specific coverage holes~\cite{kitchen2007stimulus}.
Moreover, solvers tend to produce similar solutions in successive invocations ~\cite{towards_smt_sampling}, limiting the diversity needed for comprehensive coverage exploration.
In practice, verification engineers must manually tune distribution constraints through trial and error, a labor-intensive, design-specific process that does not generalize.



Traditional verification workflows treat test selection and test generation as independent problems~\cite{guzey_dac08}. A test category is chosen, and its internal generator produces stimuli during simulation. This decoupling prevents selection from directing generation toward specific coverage targets, while generation outcomes cannot inform adaptive refinement of selection strategies, resulting in redundant stimuli and inefficient coverage convergence.

In this paper, we present \nova, a verification framework that coordinates test selection and Bayes-optimized constrained randomization to accelerate functional coverage closure.
\nova integrates seamlessly into existing CRV workflows widely adopted in industry, requiring no testbench modifications.
To the best of our knowledge, \nova is the first framework that leverages Bayesian optimization to control the solver for constrained randomization, while simultaneously integrating this process with test selection.

\nova addresses the above challenges through three integrated components.
First, \nova introduces coverage-aware test selection that clusters coverage bins based on structural and behavioral similarity at a fine granularity, enabling explicit targeting of under-covered regions from coverage behavior perspective rather than relying on feature-space diversity.
Second, \nova designs a parameterized branching strategy that exposes decision variables controlling stimuli distribution during constrained randomization, creating an optimization space that can be dynamically adjusted without testbench modifications.
Third, \nova applies Bayesian optimization to efficiently search this parameter space and identify configurations that maximize coverage growth rate, which is critical because each evaluation requires expensive full RTL simulations. Bayesian optimization is uniquely suited to such expensive black-box optimization.
\nova further leverages relevance analysis to reduce optimization dimensionality and employs a multi-stage optimization strategy to minimize optimization overhead.
Jointly, \nova coordinates test selection and constrained randomization toward under-covered regions, enabling faster and more robust coverage closure.

Our contribution can be summarized as follows:

\vspace{-2ex}

\begin{itemize}
    \item A unified verification framework that systematically coordinates coverage-aware test selection with Bayes-optimized constrained randomization to accelerate coverage closure.
    \item A coverage-aware test selection algorithm that exploits fine-grained structural and behavioral feedback, explicitly targeting under-covered regions to overcome the inefficiency of coverage-agnostic methods.
    \item A Bayes-optimized constrained randomization mechanism featuring a novel parameterized solver policy that enables dynamic control of stimuli distribution and automatically boosts coverage growth and convergence.
\end{itemize}

We conduct experiments on a range of hardware designs, from module-level IPs to system-level RISC-V processors, with constrained random verification testbenches written in SystemVerilog. Results show that NOVA can achieve up to 2.82$\times$ coverage convergence speedup compared to isolation forest\cite{nvidia_dac23} and random baseline.

%% file: sec/2_background.tex
\section{Background}

\paragraph{Constrained-Random Verification (CRV)} CRV is the most widely used industrial workflow for functional verification~\cite{spear2006systemverilog}. In CRV, SystemVerilog UVM testbench specifies constraints over input stimuli, including legality conditions and distribution constraints that encode engineers' prior knowledge.
These constraints are processed by a constraint solver, which generates concrete test vectors satisfying all requirements.
The DUT is driven with these stimuli while monitors sample selected signals at designated locations to collect functional coverage information.
Coverage is defined through \emph{covergroups}, where each covergroup contains \emph{coverpoints} observing one or more DUT signals. Each coverpoint is divided into \emph{bins}, representing distinct value ranges or scenarios, and a bin is marked covered when the corresponding condition is exercised.
Coverage closure is achieved when all bins across all coverpoints have been exercised.
In practice, closure is approximated when coverage growth plateaus or the verification time budget is reached.

\paragraph{Test selection} Test selection prioritizes tests that are most beneficial for coverage improvement.
Existing approaches employ various techniques including one-class SVM with custom kernels~\cite{guzey_dac08}, graph-based kernels for assembly tests~\cite{novelty10}, autoencoders over test parameters~\cite{vae_dvcon21}, isolation forest on runtime features~\cite{nvidia_dac23}, and LSTM-based sequence analysis~\cite{zheng2024detecting}.
These methods generally seek to identify diverse or novel tests by measuring dissimilarity in feature space or learned representations.

\paragraph{Coverage-directed test generation (CDG)} CDG steers stimuli based on coverage feedback.
Parameter-based approaches~\cite{cdg_dvcon21, graphcov} employ Bayesian optimization over a small number of tunable testbench parameters to increase hit rates of under-covered or structurally related bins.
Reinforcement learning approaches~\cite{pfeifer2020_rl_shared_memory, aref2024_rl_algorithms_design_verification, choi2021_dr_aml_dram_verification} learn stimulus policies from coverage rewards to better explore hard-to-reach scenarios.
Model-based methods like Design2Vec~\cite{design2vec} train neural network models on RTL representations to predict coverage outcomes and guide input generation through gradient.
Hardware fuzzing~\cite{rfuzz, directfuzz, difuzzrtl, hw_fuzzing, genfuzz, fuss_tecs} adapts software fuzzing techniques to hardware verification, using mutation-based exploration guided by hardware-specific coverage metrics.
While effective, these methods typically require explicit tunable parameters, custom testbench instrumentation, or specialized coverage models, making them difficult to integrate into standard CRV workflows.

%% file: sec/3_overview.tex
\section{Overview}
\input{fig/overview.tex}

The \nova framework integrates test selection and test generation into a unified flow to accelerate coverage closure. As shown in \autoref{fig:overview}, the process begins with coverage-aware test selection (\circled{1}), where the framework dynamically selects a novel test category and identifies the least-covered bins from both current and historical coverage data. This information guides the constrained randomization process (\circled{2}) to select the most appropriate pre-optimized solver parameters (tuned via Bayesian optimization) that target these coverage gaps, while the actual coverage behaviors observed from simulation runs feedback to refine subsequent test selection decisions, forming a tightly coordinated feedback loop.

To support both selection and optimization, \nova introduces an initial warm-up stage. During this phase, a fixed number of tests are executed across all categories, and the resulting coverage profiles are used to direct further test selection and cluster categories according to their coverage behavior. The clustering also leverages structural relationships such as similarity among the underlying sampled signals of individual coverpoints and vector representations derived from constituent signals in crosses. These clusters guide subsequent stages: the framework selects the cluster with the largest remaining uncovered space, and Bayesian optimization tunes solver parameters based on each target cluster's coverage characteristics. Since early verification already exhibits abundant opportunities for coverage gains even with random test selection, the warm-up incurs negligible overhead while establishing the foundation for fine-grained test selection and solver tuning in subsequent stages.

%% file: fig/overview.tex
\begin{figure}[t]
    \centering
    \includegraphics[width=0.8\linewidth]{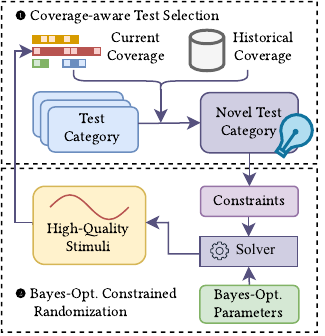}
    \caption{Overview of \nova}
    \label{fig:overview}
\end{figure}

%% file: sec/4_method.tex
\section{Test Selection Strategy}

\begin{figure}[t]
    \centering
    \includegraphics[width=\linewidth]{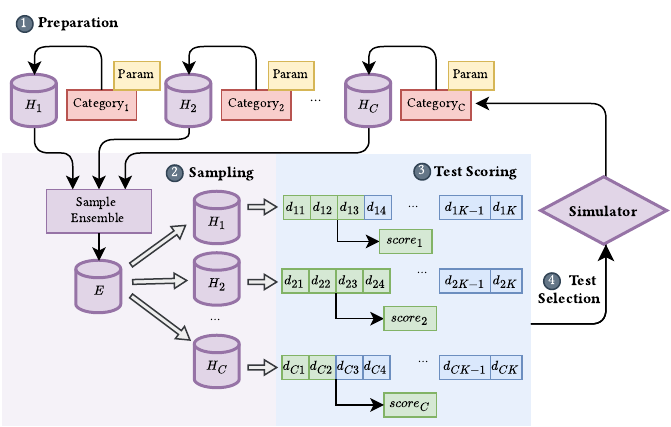}
    \caption{Test Selection Strategy}
    \label{fig:select}
\end{figure}

In this section, we address the test selection problem, which focuses on choosing the next test category to maximize coverage gain. As common bins are quickly saturated, the remaining uncovered bins become increasingly rare and unevenly distributed. Consequently, selecting an appropriate category becomes increasingly challenging, since only a small portion of categories are capable of reaching these remaining uncovered bins.

\subsection{Problem Formulation}

For clarity, we formulate the problem as follows. Given $C$ test categories and a simulation time budget $\Gamma$, the task is to select a category $c_i \in \{1, \ldots, C\}$ at each iteration $i$ and execute a test from that category. Each test produces a runtime $t_i$ and a coverage vector $cov_i \in \{0,1\}^n$, where $n$ is the total number of coverage bins. Under the constrained-random verification (CRV) setting, each test is modeled as a random draw from a distribution $D_c$ associated with its category $c$. The objective is to maximize the total number of covered bins under the given simulation time budget:

$$
\max_{\{c_i\}} \;\Bigl|~\bigcup_i cov_i~\Bigr|
\quad \text{s.t.}\quad
\sum_i t_i \le \Gamma.
$$

\subsection{Algorithm}

\begin{algorithm}[!t]
    \SetKwComment{Comment}{/* }{ */}
    \SetAlgoLined
    \caption{Test Selection}
    \label{algo:test-selection}
    \KwIn{$C$ category of tests, $G$ clusters of bins}
    \KwIn{$P$ warm-up stage simulation results}
    $H=H_1\cup H_2\cup \cdots\cup H_C \leftarrow \text{collect-coverage-vectors}(P)$ \label{line:prep} \\
    $T_{sim} = 0$ \\
    \While{$T_{sim} < \Gamma \land  \lnot (H~\text{covers}~G)$}{
        \label{line:main-loop}
        $\text{scores} \leftarrow [\cdot]$ \\
        $g \leftarrow \text{find-least-covered-cluster}(G, H)$\\
        $E \leftarrow \text{sample-ensemble}(H, g)$ \label{line:sample-ensemble-call} \\

        \For{category $c = 1\dots C$}{
            $d_c \leftarrow \text{compute-dist}(E, H_c, g)$ \label{line:scoring} \\
            $\text{mean}(\text{scores}[c] \leftarrow \text{topK}(d_c, \mathbf{R}))$ \label{line:scoring2}
        }
        $\text{choice}=\text{argmax}(\text{scores}[\cdot])$ \label{line:choice} \\
        $t, cov \leftarrow \text{simulate}(Test_{\text{choice}},g)$ \\
        $T_{sim} = T_{sim} + t$ \\
        \text{update-history-coverage}($H_{\text{choice}}$, $cov$)
    }
\end{algorithm}

The test selection flow of \nova is shown in \autoref{algo:test-selection} and \autoref{fig:select}.
It takes as inputs the test categories and a set of clustered bins to cover.
\nova leverages data collected from warm-up stage simulation as a starting point to guide future decisions. The whole flow consists of four main steps: (1) preparation, (2) sampling, (3) test scoring, and (4) selection.

\paragraph{Preparation} In \autoref{line:prep} of \autoref{algo:test-selection}, \nova organizes the warm-up simulation results into per-category historical coverage sets, providing an initial view of each category's coverage behavior. Each category is also equipped with pre-optimized parameters that provide a reasonable starting distribution for the solver to explore the constraint space; their optimization is described in \autoref{sec:bayes}.

The main loop (\autoref{line:main-loop}) repeatedly selects and executes tests based on evolving coverage information.
At each iteration, \nova identifies the least-covered cluster $g$ as the target, then evaluates all categories using a scoring mechanism that measures the distance between each category's coverage behavior and the historical coverage with respect to cluster $g$.
The category with the highest score is selected for the next test. The loop terminates once the time budget is exhausted or all bins are covered.

\begin{algorithm}[t]
    \SetKwComment{Comment}{/* }{ */}
    \SetAlgoLined
    \caption{Sample Ensemble}\label{algo:sample-ensemble}
    \KwIn{$H$ Historical coverage vectors}
    \KwIn{$g=\{b_1, b_2, \dots, b_n\}$ Bin cluster to cover}
    \KwOut{$E$ Ensemble of samples}
    $E \leftarrow \{\}$ \\
    \For{$e=1\dots \textbf{Ensemble\_Size}$}{
        $S \leftarrow \{\}$ \\
        \For{$b_i$ in $g$}{ 
            $S \leftarrow S \cup \text{uniform-sample}(\{h \; | \; h \in H \; \text{covers} \; b_i\}, \textbf{Samples\_Per\_Bin})$  \label{line:sample-detail}
        }
        $E \leftarrow E \cup \{\text{uniform-sample}(S, \textbf{Sample\_Size})$\}
    }
\end{algorithm}

\paragraph{Sampling}
In \autoref{line:sample-ensemble-call}, \nova constructs a balanced sample ensemble $E$ using the procedure defined in \autoref{algo:sample-ensemble}. The distribution of historical coverage vectors $H$ is highly skewed: bins that are frequently exercised appear in many vectors, while rare bins occur in only a small subset of $H$. Direct sampling would therefore preserve this imbalance.
To counteract this skew, the routine first performs bin-level uniform sampling: for each bin in the target cluster $g$, it extracts an equal number of coverage vectors from the subset of $H$ that cover that bin, forming an intermediate set $S$. This stratification equalizes per-bin representation regardless of each bin's empirical frequency. A second sampling step then uniformly draws a fixed number of vectors from $S$ to control the sample size and avoid distortion caused by overly large intermediate sets. To enhance robustness, this two-stage process is repeated multiple times, and the resulting representative sets are aggregated to form the final ensemble $E$.

\paragraph{Test Scoring} In \autoref{line:scoring}, \nova measures the novelty of each category $c$ by comparing its historical coverage vectors $H_c$ against the sample ensemble $E$ targeting bin cluster $g$. Ideally, for every historical vector $h_{ck} \in H_c$, one would compute its distance $d_{ck}$ to all other historical vectors in $H$, forming a vector $d_c=\{d_{ck}\}$ to represent the distance in coverage behavior between category $c$ and historical tests, but this requires an $O(|H|^2)$ pairwise comparison at each iteration.
To avoid this cost, \nova adopts the ensemble-based scoring strategy of LeSiNN~\cite{lesinn}, which approximates novelty using a collection of representative subsets $S_1, S_2, \dots, S_{\text{|E|}}$ generated by \autoref{algo:sample-ensemble}. Each $S_e$ is a bin-balanced subset of $H$, capturing diverse behaviors while being much smaller than the full historical set. Using these subsets as surrogates, the distance of a vector $h_{ck} \in H_c$ to the ensemble is defined as
\begin{equation} \label{eqn:dist-func}
    d_{ck} = \sum_{i=1}^{\text{Ensemble\_Size}} nn\_dist(h_{ck}, S_e, g),
\end{equation}
where $nn\_dist(h_{ck}, S_e, g)$ denotes the nearest-neighbor distance from $h_{ci}$ to subset $S_e$, with coverage vector respect to bin cluster $g$. The number of vectors in $E$ is a constant much smaller than $|H|$. This approximation reduces the scoring complexity to $O(|H|)$ while preserving the ability to capture meaningful behavioral differences across categories.

To prevent rare scenarios from being averaged out by common ones, \nova focuses on only the most distinctive samples when computing category scores in \autoref{line:scoring2}. Specifically, the scoring step uses a hyper-parameter $\mathbf{R}$ to control the fraction of samples considered: for each test category, \nova computes the score $\text{scores}[c]$ as the mean distance of its top $\mathbf{R}\%$ most distant samples in $d_c$.
The parameter $\mathbf{R}$ should be set in proportion to the likelihood that a rare scenario happens, which is usually less than 1\%.

\paragraph{Test Selection}
In \autoref{line:choice}, \nova selects the test category with the highest score and then executes a test from it using its pre-trained solver parameters for the target bin cluster $g$, thereby generating new simulation coverage. After the simulation completes, \nova updates the historical coverage set of the selected category with the newly obtained vectors.

\section{Bayes-Optimized Constrained Randomization}
\label{sec:bayes}

In this section, we introduce a Bayes-optimized constrained randomization test-generation method that adaptively shapes stimulus distributions during CRV. By parameterizing the solver, defining a continuous cluster-level coverage objective, and tuning distribution parameters through Bayesian optimization, this method guides test generation toward stimuli that more effectively exercise coverage-relevant behaviors.

\subsection{Parameterized Solver}

\begin{figure}
    \centering
    \includegraphics[width=\linewidth]{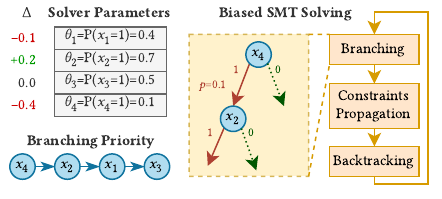}
    \caption{Parameterized Solver}
    \label{fig:solver}
\end{figure}

In CRV, the distribution of generated stimuli is traditionally controlled only via manually crafted dist constraints in SystemVerilog, which is inflexible and labor-intensive.
In \nova, we introduce a parameterized solver strategy, as illustrated in \autoref{fig:solver}, where each branching decision is associated with a tunable parameter $\theta_i \in (0,1)$, representing the probability of assigning $x_i{=}1$.
The solver arranges branching priorities by ranking variables in $|0.5-\theta_i|$, so variables whose probabilities deviate more from $0.5$ are branched on earlier, exerting greater influence on the resulting stimulus distribution.
For unconstrained variables, the strategy induces a Bernoulli distribution, while for constrained variables, it biases assignments toward 0 or 1 within the valid solution space.
By tuning the parameter vector $\theta$, we continuously shape the test distribution $D_c^\theta$ for each category $c$, yielding a smooth optimization space that is readily amenable to automated search.

\subsection{Objective Function Design}

A central challenge is to define a meaningful and optimizable objective. Existing approaches typically maximize either (i) the number of overall covered bins or (ii) the probability of hitting a specific target bin. However, these formulations exhibit severe drawbacks. First, for overall coverage, the heterogeneity of bins induces a discontinuous objective surface, impeding gradient-free optimization methods. Second, for single-bin targeting, the objective degenerates into a binary 0/1 metric, which provides no intermediate feedback and thus offers little guidance for optimization.

To overcome these limitations, we propose a cluster-based sample-rate metric. Specifically, bins are grouped into semantically related clusters $g$. For each cluster, we compute its empirical sample rate, denoted $rate_{g}^\theta$, defined as the number of times bins in the cluster are hit during a given sampling window. Assuming a uniform distribution across the $|g|$ bins of the cluster, the expected number of unique covered bins is

\begin{equation}
    \mathbb{E}_{g} = |g|\cdot \left(1-\left(1-\frac{1}{|g|}\right)^{rate_{g}^\theta}\right)
\end{equation}
which directly follows from the expectation of Bernoulli trials for repeated sampling. This formulation transforms the discrete hit-or-miss outcome into a continuous objective, while still reflecting the coverage potential of a cluster. 

The overall optimization problem to maximize the expected coverage $\mathbb{E}_{g}$ of target cluster, where solver branch probability parameters $\theta$ are variables. To estimate $\mathbb{E}_{g}$, we sample runtime statistics under a given configuration of $\theta$, measure and calculate cluster-wise hit frequencies from sampled simulation times and coverage vectors, then compute the empirical expectation. This black-box objective is later optimized via Bayesian optimization with Sparse Axis-Aligned Subspaces (SAASBO)~\cite{eriksson2021high}, which is well suited for high-dimensional space and expensive objectives.

\subsection{Scalability Enhancements}

\paragraph{Multi-Stage Optimization}
While SAASBO is well-suited to expensive objectives, its computational complexity increases with both the number of parameters and the number of coverage clusters. To address this challenge, we adopt a two-stage procedure. First, 
\textit{Global optimization} maximizes the aggregated expectation across all clusters to quickly locate promising parameter regions. Second, \textit{Fine-tuning} refines results by sequentially optimizing expectations for individual clusters, thereby reducing dimensionality and computational overhead.
Since early verification already exhibits abundant opportunities for coverage gains—even with random test selection—the warm-up incurs negligible overhead while establishing the foundation for fine-grained test selection and solver tuning in subsequent stages.
\paragraph{Relevance-Guided Parameter Reduction}
To further improve optimization efficiency, we introduce a relevance-guided parameter reduction technique based on \textit{SHAP analysis}~\cite{shapley1951notes,NIPS2017_7062}. Specifically, we identify random variables that exhibit strong correlations with the sampled signals of coverpoints and exclude them from the parameterized solver. By removing these dominant variables, the remaining sampling process becomes closer to our uniform distribution assumption within each cluster. At the same time, this exclusion substantially reduces the dimensionality of the optimization space. Consequently, the optimization avoids concentrating on specific bins, enabling a more balanced and generalizable process aligned with our goal of accelerating coverage closure.

%% file: sec/5_eval.tex
\section{Experiments}

\subsection{Setup}

We evaluate \nova using five hardware designs spanning IP-level components to system-level processors: AON Timer and UART controller from OpenTitan~\cite{opentitan}, AXI4~\cite{mbits_mirafra_axi4_avip_2025}, the Ibex RISC-V core~\cite{lowrisc_ibex_2025}, and the CV32E40P core from OpenHW Group~\cite{cv32e40p}. Table~\ref{tab:benchmarks} summarizes the key statistics of these designs and their verification configurations. All IP-level benchmarks use constrained-random verification with user-defined functional coverage models from their respective repositories. For processor cores, we adopt CRV-generated assembly tests from riscv-dv~\cite{noauthor_chipsallianceriscv-dv_2025}, which are compiled, executed on the DUT, and evaluated against functional coverage models.
We implement \nova on top of VeriSim~\cite{verisim_en}, a commercial event-driven RTL simulator. \nova does not rely on VeriSim-specific features and can be easily adapted to other platforms such as Verilator~\cite{Verilator} or Synopsys VCS~\cite{vcs} by modifying the solver interface.

Our evaluation consists of three parts: (1) overall performance comparing \nova against random selection and isolation-forest-based selection~\cite{nvidia_dac23} across all benchmarks, (2) analysis of Bayesian optimization on UART and AXI4 against human-engineered constraints, (3) case study of UART including design choice ablation, runtime breakdown analyzing the overhead of Bayesian Optimization and test selection, and comparison with state-of-the-art LSTM autoencoder-based method~\cite{zheng2024detecting}. For fair comparison, all methods use the same initial pool of executed tests during the warm-up stage, where we run 10 tests from each category, and results are reported only after warm-up completion.

\vspace{-1ex}

\subsection{Overall Performance}

As shown in \autoref{fig:overall}, \nova demonstrates consistent simulation time reduction across benchmarks, achieving speedups ranging from $1.43\times$ to $2.82\times$ (average $1.86\times$) over isolation forest~\cite{nvidia_dac23} and random selection baselines. While baselines use human-engineered randomization constraints, \nova leverages Bayes-optimized constrained randomization. The AON Timer benchmark achieves 98.3\% coverage with $2.03\times$ speedup, demonstrating that high coverage and efficiency are not mutually exclusive goals. The UART benchmark reaches 96.9\% coverage with $1.43\times$ speedup.
The Ibex results highlight \nova's ability to avoid the pitfalls of naive optimization: while the isolation forest baseline stagnates by over-selecting interrupt tests with long runtimes but with minimal coverage contribution, \nova efficiently identifies coverage-critical categories, achieving an impressive $2.82\times$ speedup. For CV32E40P, both approaches identify valuable test categories, but \nova's optimized randomization enables reaching 79.5\% coverage with $1.50\times$ speedup. 
For AXI4, NOVA achieves a $1.84\times$ speedup, even though the reported 74.2\% coverage is deflated due to the coverage calculation including unreachable and invalid states intrinsic in this benchmark. The testbench already incorporates extensive manually crafted test cases, resulting in a long warm-up stage but with limited effectiveness in closing coverage gaps. NOVA's systematic identification of high-contribution tests enables substantially faster convergence.

\input{tab/setup.tex}
\input{fig/overall.tex}

\subsection{Analysis of Bayesian Optimization}
\input{fig/bayes.tex}

We compare Bayes-optimized constrained randomization against the human-engineered randomization constraints provided in the UART~\cite{opentitan} and AXI4~\cite{mbits_mirafra_axi4_avip_2025} testbenches. These human-engineered constraints specify manually designed distribution settings that verification engineers consider effective for achieving coverage closure. For fairness, the comparison is conducted under the same test category and target cluster.
As shown in \autoref{fig:bayes-vs-stock}, Bayes-optimized constrained randomization achieves the highest coverage of 88.2\% on UART and 73.6\% on AXI4, corresponding to 2.93$\times$ and 1.45$\times$ speedup, respectively, and consistently outperforms the manually crafted constraints. By automatically exploring the constraint space, \nova identifies configurations that generate more effective stimuli and yield higher coverage. In contrast, manually designed distributions can be suboptimal and may fail to capture the scenarios most relevant for coverage.

\subsection{Case Study on UART}

We conduct a comprehensive case study on UART. As a standard serial communication protocol with moderate complexity, UART provides an ideal controlled environment for detailed experimentation while maintaining practical relevance to real-world verification scenarios. In this case study, we analyze four aspects of our framework: the impact of reclustering in warm-up stage, the effect of excluding dominant variables during solver parameterization, the runtime breakdown across different components, and the comparison against state-of-the-art learning-based methods.


\input{fig/ablation_study.tex}

\paragraph{Impact of Reclustering.}
Figure~\ref{fig:ablation-reclustering} compares \nova with and without the reclustering mechanism performed during the warm-up stage. By reorganizing coverage bins based on early simulation behavior rather than coarse predefined groups, the refined clusters provide more informative guidance for subsequent selection. Without reclustering, coverage plateaus at 96.24\%, whereas behavior-driven clusters enable continued progress and reach 96.81\%. This improvement stems from balancing granularity: coarse coverage groups mix easy-to-cover bins and hard ones, while single-bin granularity is too fine to exploit inter-bin similarities, making the reclustered structure particularly effective in the tail phase of verification.

\paragraph{Impact of Variable Exclusion.}
During Bayesian optimization, we exclude dominant random variables to prevent highly skewed stimuli distributions. Figure~\ref{fig:ablation-excluded-variables} compares two strategies: (1) including all variables, and (2) excluding the dominant ones. When all variables are included, bias toward the most influential variables concentrates coverage on a few coverpoints, with the curve stagnating at 19.3\%. Excluding these variables preserves uniformity, enabling more balanced exploration, achieving 89.70\% coverage (4.64$\times$ improvement).

\paragraph{Runtime Breakdown.}
We profile the runtime of different components of \nova on the UART benchmark, measuring time to reach maximum coverage (Figure~\ref{fig:runtime-breakdown}). Simulation dominates at 93.79\% of total runtime, while Bayesian optimization, including trial simulations, accounts for 6.15\%, and test selection contributes only 0.07\%. The cost of Bayesian optimization is a one-time overhead that can be amortized across subsequent regression runs, as the optimized weights can be reused as the DUT evolves. Despite this small overhead, \nova achieves substantial efficiency gains.


\paragraph{Comparison with State-of-the-Art}
The state-of-the-art LSTM-based approach~\cite{zheng2024detecting} predicts the novelty of each test category by training an LSTM combined with an autoencoder to estimate a novelty score, and then selects tests based on this predicted score. Compared to this method on the UART benchmark, \nova achieves higher coverage at 96.8\% and converges 1.35$\times$ faster. The key advantage of \nova stems from coverage-aware test selection: while the LSTM approach relies on static test features and the predicted novelty score without considering actual coverage outcomes, leading to plateaus as shown in Figure~\ref{fig:ablation-sota}, \nova dynamically evaluates the coverage contribution of each category and identifies the most effective tests, avoiding selections that appear promising from features alone but contribute minimally to coverage goals.

%% file: tab/setup.tex
\begin{table}[t]
    \centering
    \caption{RTL Designs in Evaluation}
    \label{tab:benchmarks}
    \SetTblrInner{rowsep=0.5pt}
    \begin{tblr}{c|ccc}
        \hline[1pt]
        \SetRow{font=\bfseries}
        Design &  Test categories & Test bins & Parameters \\
        \hline
         AON Timer  &  12  &  2048  & / \\
         UART       &  12  &  4417  & 729 \\
         AXI4   &  62  &  2272 & 106 \\
         Ibex       &  57  &  11396 & 476 \\
         CV32E40P   &  46  &  12752 & 416 \\
        \hline[1pt]
    \end{tblr}
\end{table}

%% file: fig/overall.tex
\begin{figure}
    \centering
    {
    \scriptsize
    \noindent
    \textcolor{myorange}{\shortline} NOVA \quad
    \textcolor{myblue}{\shortline} Isolation forest \quad
    \textcolor{myred}{\shortline} Random selection
    }

    \def\tsfigurewidth{0.23\textwidth}
    \begin{subfigure}[t]{\tsfigurewidth}
        \centering
        \includegraphics[width=\textwidth]{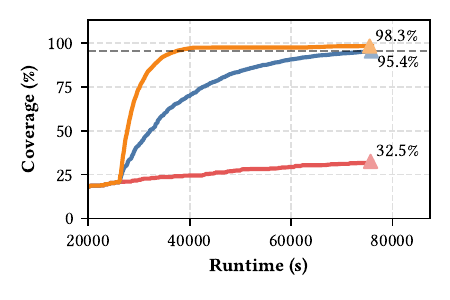}
        \caption{AON Timer}
        \label{overall:timer}
    \end{subfigure}
    \begin{subfigure}[t]{\tsfigurewidth}
        \centering
        \includegraphics[width=\linewidth]{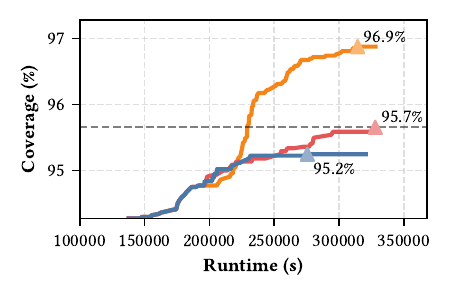}
        \caption{UART}
        \label{overall:uart}
    \end{subfigure}
    \begin{subfigure}[t]{\tsfigurewidth}
        \centering
        \includegraphics[width=\textwidth]{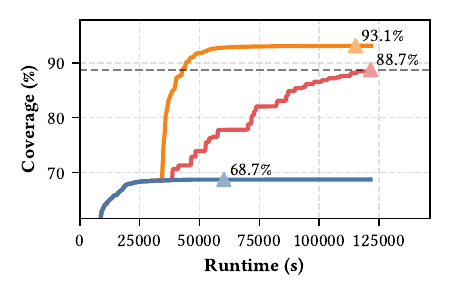}
        \caption{Ibex}
        \label{overall:ibex}
    \end{subfigure}
    \begin{subfigure}[t]{\tsfigurewidth}
        \centering
        \includegraphics[width=\textwidth]{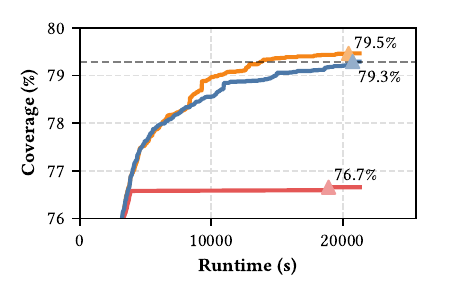}
        \caption{CV32E40P}
        \label{overall:corev}
    \end{subfigure}
    \begin{subfigure}[t]{\tsfigurewidth}
        \centering
        \includegraphics[width=\textwidth]{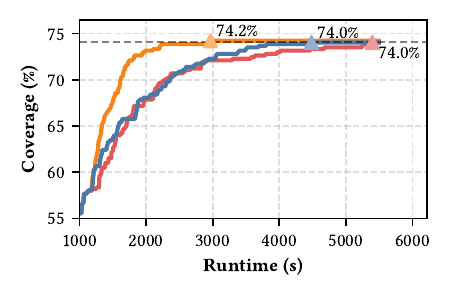}
        \caption{AXI4}
        \label{overall:axi}
    \end{subfigure}
    \begin{subfigure}[t]{\tsfigurewidth}

        \raisebox{1.1\height}{
        \hspace{1.5em}

        \resizebox{0.8\linewidth}{!}{
            \SetTblrInner{rowsep=1pt}
            \begin{tblr}{
                |cc|
            }
                \hline
                \textbf{Benchmark} & \textbf{Speedup} \\ 
                \hline
                AON Timer & $2.03\times$ \\
                UART & $1.43\times$ \\
                Ibex & $2.82\times$ \\
                CV32E40P & $1.50\times$ \\
                AXI4 & $1.84\times$ \\
                \hline
            \end{tblr}
        }
        }

        \caption{Speedup Summary}
        \label{overall:summary}
    \end{subfigure}

    \caption{Overall speedups of {\nova}~over {isolation forest}~and {random selection} baselines across all benchmarks.}
    \label{fig:overall}
\end{figure}

%% file: fig/bayes.tex
\begin{figure}[t]
    \centering
    {
        \scriptsize
        \noindent
        \textcolor{myorange}{\shortline} Bayes-optimized constrained randomization \quad
        \textcolor{myblue}{\shortline} Human-engineered randomization
    }

    \vspace{3pt}

    \def\tsfigurewidth{0.23\textwidth}
    \begin{subfigure}[t]{\tsfigurewidth}
        \centering
        \includegraphics[width=\linewidth]{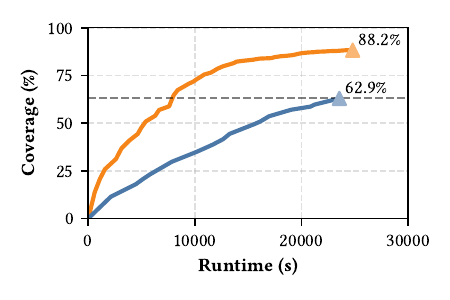}
        \caption{UART}
        \label{fig:bayes-uart}
    \end{subfigure}
    \hfill
    \begin{subfigure}[t]{\tsfigurewidth}
        \centering
        \includegraphics[width=\linewidth]{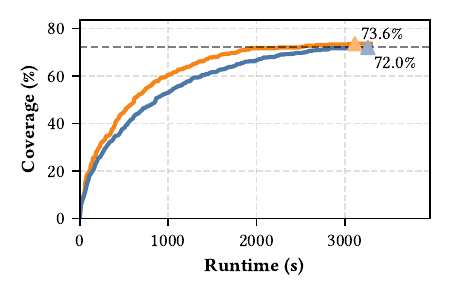}
        \caption{AXI4}
        \label{fig:bayes-axi}
    \end{subfigure}

    \caption{{Bayes-optimized constrained randomization} consistently outperforms {human-engineered randomization} across both UART and AXI4 benchmarks.}
    \label{fig:bayes-vs-stock}
\end{figure}

%% file: fig/ablation_study.tex
\begin{figure}
    \centering

    {
        \scriptsize
        \noindent
        \textcolor{myorange}{\shortline} Full NOVA \quad
        \textcolor{myblue}{\shortline} Ablated NOVA \quad
        \textcolor{mygreen}{\shortline} SOTA \quad
    }

    \def\tsfigurewidth{0.235\textwidth}
    \begin{subfigure}[t]{\tsfigurewidth}
        \centering
        \includegraphics[width=\linewidth]{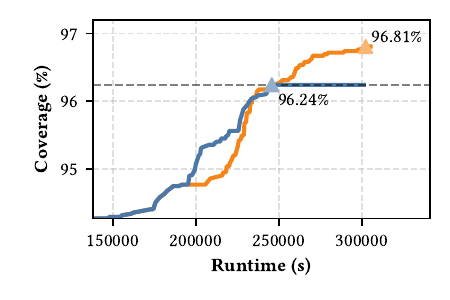}
        \caption{Re-clustering of coverage bins}
        \label{fig:ablation-reclustering}
    \end{subfigure}
    \begin{subfigure}[t]{\tsfigurewidth}
        \centering
        \includegraphics[width=\linewidth]{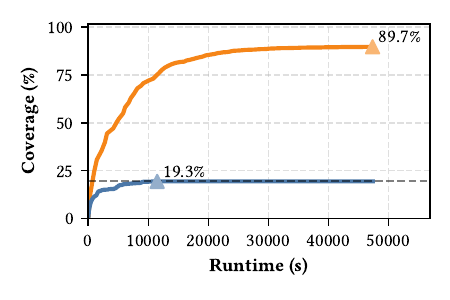}
        \caption{Excluded variables}
        \label{fig:ablation-excluded-variables}
    \end{subfigure}
    \begin{subfigure}[t]{\tsfigurewidth}
        \centering
        \includegraphics[width=\linewidth]{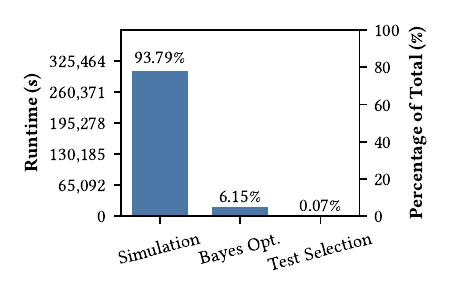}
        \caption{Runtime breakdown}
        \label{fig:runtime-breakdown}
    \end{subfigure}
    \begin{subfigure}[t]{\tsfigurewidth}
        \centering
        \includegraphics[width=\linewidth]{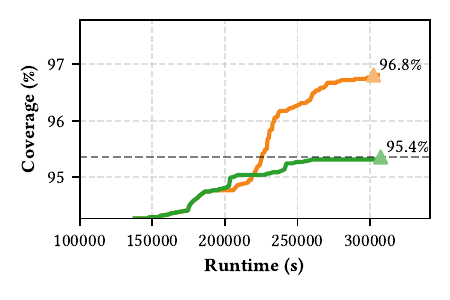}
        \caption{Comparison with SOTA}
        \label{fig:ablation-sota}
    \end{subfigure}
    
    \caption{Design choice ablation, runtime breakdown, and comparison with SOTA LSTM autoencoder on UART.}
\end{figure}

%% file: sec/7_conclusion.tex
\section{Conclusion}

In this work, we presented \nova, a framework that accelerates coverage closure by coordinating test selection with Bayes-optimized constrained randomization. By leveraging fine-grained coverage features, \nova filters out redundant tests and focuses simulation resources on those most likely to contribute new coverage. The parameterized solver exposes tunable decision strategies, which Bayesian optimization automatically adjusts to maximize coverage growth rate without relying on manual heuristics. Experimental results across multiple RTL designs and verification environments demonstrate that \nova achieves up to 2.82$\times$ faster coverage convergence, highlighting the effectiveness of coordinated, data-driven verification strategies.